\documentclass[journal]{IEEEtran}
\usepackage{cite}
\usepackage{amssymb}
\usepackage[cmex10]{amsmath}
\usepackage{graphicx}
\usepackage{cases}
\usepackage{array}
\usepackage{color}

\newcommand{\figref}[1]{Fig.\hspace{0.03in}\ref{#1}}
\newcommand{\Eqref}[1]{Eq.\hspace{0.03in}(\ref{#1})}
\newcommand{\insref}[2]{\figref{#1}\hspace{0.01in}({#2})}

\begin{document}
\title{Memory Models of Adaptive Behaviour}
\author{Fabio~Lorenzo~Traversa, Yuriy~V.~Pershin,~\IEEEmembership{Member,~IEEE,}~and~Massimiliano~Di~Ventra
\thanks{F. L. Traversa is with the Department of Electronic Engineering, Universitat Aut\`{o}noma de Barcelona, 08130, Spain and with the Department of Physics, University of California, San Diego, La Jolla, California 92093-0319, USA, e-mail: fabiolorenzo.traversa@uab.es}
\thanks{Y. V. Pershin is with the Department of Physics and Astronomy, University of South Carolina, Columbia, South Carolina 29208, USA, e-mail: pershin@physics.sc.edu}
\thanks{M. Di Ventra is with the Department of Physics, University of California, San Diego, La Jolla, California 92093-0319, USA,  e-mail: diventra@physics.ucsd.edu}}
\maketitle

\begin{abstract}
Adaptive response to a varying environment is a common feature of biological organisms. Reproducing
such features in electronic systems and circuits is of great importance for a variety of applications.
Here, we consider memory models inspired by an intriguing ability of slime molds to both memorize the period of temperature and humidity variations, and anticipate the next variations to come, when appropriately trained. Effective circuit models of such behavior are designed using {\it i}) a set of LC-contours with memristive damping, and {\it ii}) a single memcapacitive system-based adaptive contour with memristive damping. We consider these two approaches in detail by comparing their results and predictions. Finally, possible biological experiments that would discriminate between the models are discussed. In this work, we also introduce an effective description of certain memory circuit elements.
\end{abstract}

\begin{IEEEkeywords}
Memory, Memristor, Memcapacitive system, Adaptive Frequency, Synchronization, Learning, Amoeba, Dynamical Systems
\end{IEEEkeywords}

%%%%%%%%%%%%%%%%%%%%%%%%%%%%%%%%%%%%%%%%%%%%%%%%%%%%%%%%%%%%%
%%%%%%%%%%%%%%%%%%%%%%%%%%%%%%%%%%%%%%%%%%%%%%%%%%%%%%%%%%%%%
\section{Introduction}
%%%%%%%%%%%%%%%%%%%%%%%%%%%%%%%%%%%%%%%%%%%%%%%%%%%%%%%%%%%%%
%%%%%%%%%%%%%%%%%%%%%%%%%%%%%%%%%%%%%%%%%%%%%%%%%%%%%%%%%%%%%

Adaptive behavior is common in Nature and may find useful applications if implemented in electronics \cite{our_chapter}.
Generally, adaptive behavior is related to memory -- the ability to store and retrieve relevant information. Recently,
it was shown experimentally that an amoeboid organism -- the slime mold {\it Physarum polycephalum} -- employs both internal \cite{Saigusa_2008} and external \cite{reid12a} "memories" to keep information about past events and utilize this information in future responses. While the external memory of slime molds is organized similarly to that of ant colonies \cite{Jackson06a,book_ants} (a moving plasmodium
leaves behind a thick layer of an extracellular slime whose sense determines its future behavioral response \cite{reid12a}), the
origin of the internal memory is still unclear. In addition to the memory feature, it was also demonstrated that slime molds can solve mazes \cite{Nakagaki00a,Nakagaki01a} and other shortest-path problems \cite{Nakagaki04a,Nakagaki04b,Nakagaki07a}. The memory, problem solving and adaptive abilities of {\it Physarum polycephalum} are surprising since slime molds are unicellular organisms without a neural system.

In experiments on the internal memory \cite{Saigusa_2008}, the locomotion speed of slime molds was studied as a function of environmental temperature and humidity. It was observed that slime molds recognize periodic changes in the environment, memorize their periods, and adjust their future behavior based on the memorized information. Taking into account the available experimental information, several models can be put forward to describe the slime molds' memory and response.
In this paper, we will discuss two electronic schemes for memory organization in simple organisms inspired by the intriguing internal memory abilities of  {\it Physarum polycephalum} \cite{Saigusa_2008}. These electronic models
of the slime mold adaptation serve two important purposes: {\it i}) they are interesting adaptive circuits that
can find applications in diverse areas of electronics where adaptation to incoming signals is of interest, e.g., in pattern recognition; {\it ii}) they serve as electronic tools to predict possible biological responses that can then be tested in experiments on actual organisms.

First of all, we would like to summarize the available knowledge on behavioral abilities of amoebas (in the context of the internal memory). Together with available experimental information, we also formulate assumptions (again, in the context of the internal memory) that will help in the discussion and formulation of memory models. Below, we will then discuss experimental facts ({\bf F})
(essentially, from Ref. \cite{Saigusa_2008}), more probable assumptions ({\bf A}), and speculations/predictions ({\bf S}). Let us consider these in the order from the most reliable to the least reliable.

{\bf F1:} The locomotion speed depends on environmental conditions determined by two variables -- temperature and humidity. Favorable conditions for faster locomotion are higher temperatures and humidities. {\bf F2:} Subjected to specific periodic environmental changes, amoebas "memorize" the period and anticipate new changes to come \cite{Saigusa_2008}. Periodic slow-downs are observed after both a sequence of three training pulses and a single pulse. {\bf F3:} Amoebas "memorize" a wide range of periods. {\bf F4:} The memory about the period decays in time (longer time intervals between the training sequence and testing pulse decrease the memory response \cite{Saigusa_2008}). {\bf F5:} On average, untrained amoebas do not slow-down after a single pulse.

\begin{figure*}
\centerline{
\includegraphics[width=1.4\columnwidth]{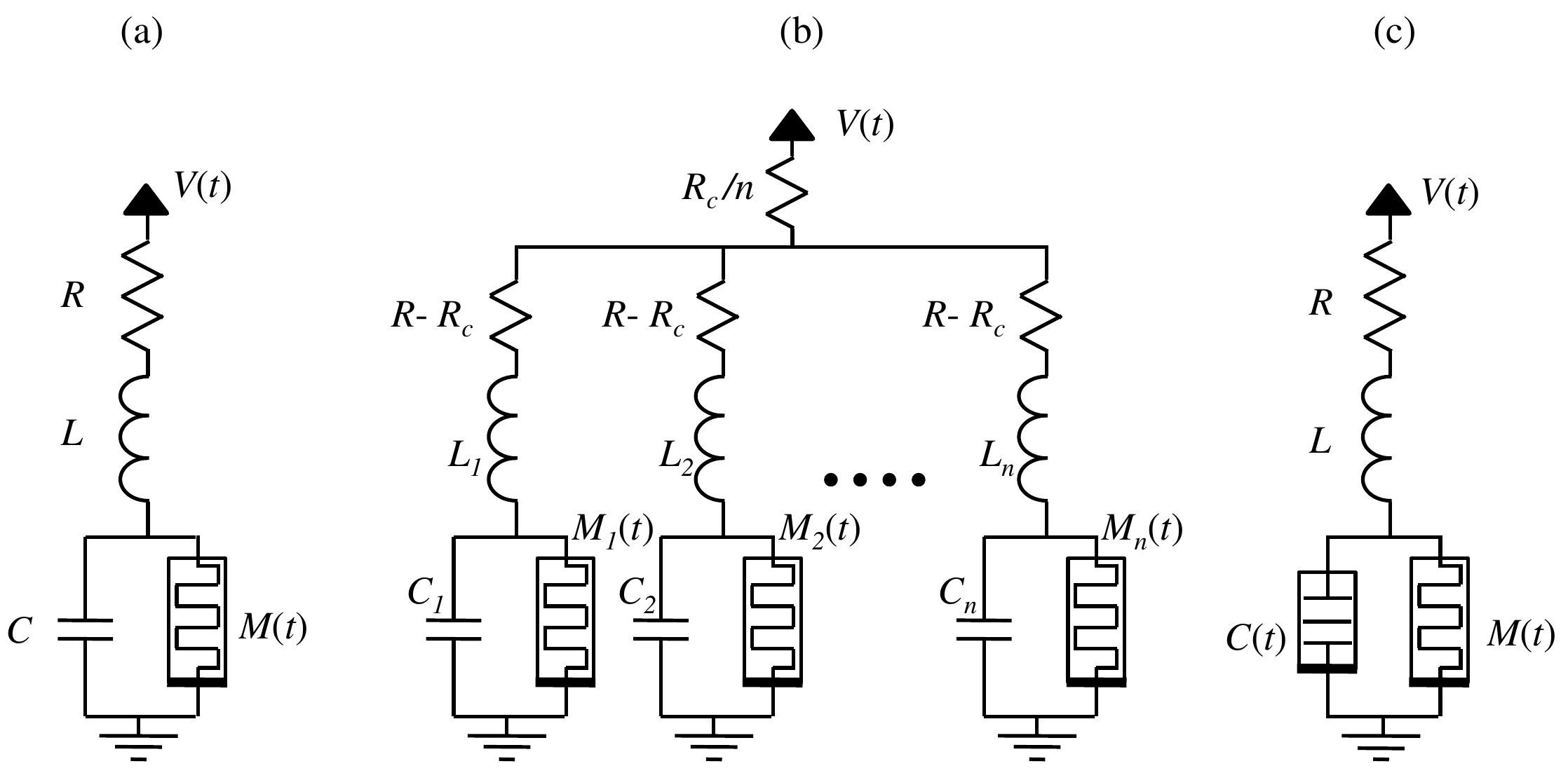}}
\caption{\label{Circuits}Equivalent electronic circuit models of amoeba\rq{s} learning: (a) single LC-contour with memristive damping from Ref. \cite{Amoeba_yuriy}, (b) $n$ coupled LC-contours with memristive damping, and (c) memcapacitive system-based adaptive contour with memristive damping.
In (b), the oscillators have different resonant frequencies covering a frequency spectrum from $\omega_1\approx (L_1C_1)^{-1/2}$ to $\omega_n\approx (L_nC_n)^{-1/2}$. Moreover, $R_c$ is a coupling resistance that reduces $R$ from (a) to $R-R_c$ in (b) to ensure the same response when a single pulse is applied. In (c), the use of memcapacitive system $C(t)$ results in a contour with adaptive frequency.}
\end{figure*}

{\bf A1:} The transition from favorable to unfavorable environmental conditions can be parameterized by a single parameter selected in a such way that the {\it equilibrium} locomotion speed changes monotonously with this parameter. {\bf A2:} There is an internal mechanism for memory decay independent of the instantaneous environmental conditions. However, the past memory can be reinforced by an external stimulus of the same period. {\bf A3:} The memory decay rate can be considered as frequency-independent. {\bf A4:} In single pulse experiments, untrained amoebas do not exhibit any slow-downs. {\bf A5:} There exists a maximum possible speed for amoebas' locomotion.

{\bf S1:} There is a threshold for learning. In other words, periodic but relatively small changes in environmental conditions cannot be memorized. {\bf S2:} If amoebas were trained to two different frequencies both frequencies would be revealed in the response. {\bf S3:} In an {\it inverse} experiment, namely, if unfavorable conditions were interrupted by periodic pulses of favorable conditions amoebas would memorize the period of pulses and demonstrate spontaneous acceleration in the anticipation of next pulses to come.

The current amount of available experimental data is not sufficient to support or reject assumptions {\bf A}s and speculations {\bf S}s. However, it is expected that a viable electronic model of behavioral abilities of amoebas must satisfy all the facts {\bf F}s, and most (if not all) of the assumptions {\bf A}s.

Previously, two possible schemes of adaptive behavior were suggested by two of us (YP and MD)\cite{Amoeba_yuriy}.  The first approach employs an array of LC-contours with additional damping elements (resistor and memristive system \cite{chua76a} in each contour). The second is a single adaptive contour involving memcapacitive and/or meminductive systems \cite{diventra09a,diventra09b}. In Ref. \cite{Amoeba_yuriy}, the response of only a single LC-contour with memristive damping element was explicitly considered, and the second scheme was only suggested at the end of
that paper.

The purpose of this publication is to develop better models of amoeba's learning following the two general schemes initially suggested in Ref. \cite{Amoeba_yuriy}. Unlike the previous publication \cite{Amoeba_yuriy}, in this one (see Sec. \ref{sec3}) we consider a {\it collective} response of an array of LC-contours covering an interval of frequencies. This consideration is based on a different model of memristive system resulting in a closer similarity with experimental observations. Several possible realizations of the adaptable LC-contour are considered in Sec. \ref{sec4}. In this Section, we also introduce an effective description of certain memcapacitive systems (see Sec. \ref{effect}). We conclude in Sec. \ref{sec5} with a brief summary, final remarks and suggestions of additional experiments that would help distinguishing between the suggested models, and thus further clarify the origins of learning and adaptive behavior in simple organisms.

%%%%%%%%%%%%%%%%%%%%%%%%%%%%%%%%%%%%%%%%%%%%%%%%%%%%%%%%%%%%%
%%%%%%%%%%%%%%%%%%%%%%%%%%%%%%%%%%%%%%%%%%%%%%%%%%%%%%%%%%%%%
\section{Equivalent Circuit Description} \label{sec2}
%%%%%%%%%%%%%%%%%%%%%%%%%%%%%%%%%%%%%%%%%%%%%%%%%%%%%%%%%%%%%
%%%%%%%%%%%%%%%%%%%%%%%%%%%%%%%%%%%%%%%%%%%%%%%%%%%%%%%%%%%%%

\figref{Circuits} presents three equivalent circuit models of adaptive behavior that are referred to in this paper.  All of the circuits are based on damped LC-contour(s) associated with  biological oscillators in {\it Physarum polycephalum}. Possible relations between the circuits' components and biological processes are the same as in Ref. \cite{Amoeba_yuriy}. The memristive systems $M$ and memcapacitive system $C$ summarize the relevant memory mechanisms in slime molds while the external voltage $V(t)$ reproduces the external stimuli such as temperature and humidity variations. The amoeba's response (the speed of locomotion \cite{Saigusa_2008}) is related to the voltage across the (mem)capacitor(s). The first circuit, \insref{Circuits}{a}, represents slime molds that can "memorize" frequencies only in a narrow fixed range. The second one, \insref{Circuits}{b}, allows learning of more than one frequency in a wide range. The third circuit, \insref{Circuits}{c}, provides learning of a single frequency but in a wide range.

\begin{figure}%\vspace{0.14cm}
\centerline{
\includegraphics[width=0.6\columnwidth]{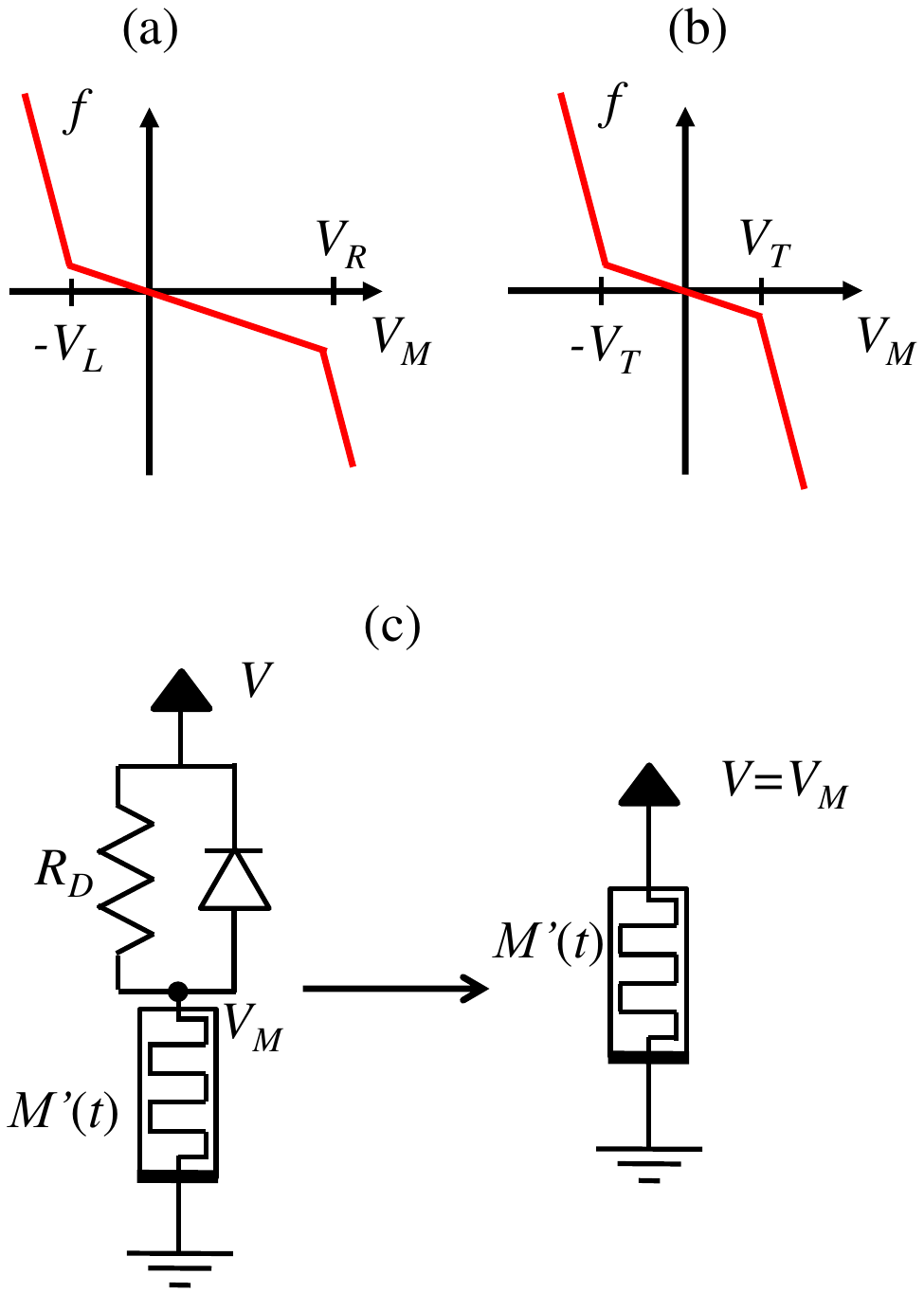}}
\caption{\label{Equivalence} (a) Schematics of the switching function $f(V)$ given by Eq. (\ref{eq3}). Asymmetry in switching thresholds can be obtained by combining a memristive system with symmetric switching function (b) with non-linear elements, such as a diode (c). The total scheme can be considered as a single effective memristive device.}
\end{figure}

The response of all circuits presented in \figref{Circuits} depends crucially on the type of memristive system used (the model of memcapacitive system $C(t)$ for \insref{Circuits}{c} is discussed in Sec. \ref{sec4}). In this work, we will use a voltage-controlled memristive device with asymmetric voltage thresholds. Mathematically, such a model is described by the following set of equations
\begin{align}
&I= x^{-1}V_M, \label{eq1} \\
&\dot x= f(V_M)[\theta(V_M)\theta(x-M_1){+}\theta(-V_M)\theta(M_2-x)] , \label{eq2}  \\
&f(V)= {-}\beta V{+}\frac{\beta-\alpha}{2}(|V{+}V_L|{-}|V{-}V_R|{+}V_R{-}V_L)\label{eq3}
\end{align}
where $I$ and $V_M$ are the current through and the voltage drop on the device, respectively, and
$x$ is the internal state variable playing the role of memristance, $M\equiv x$,
$\theta(\cdot)$ is the step function, $\alpha$  and $\beta$
characterize the rate of memristance change at lower ($|V_M|$ is below threshold) and higher ($|V_M|$ is above threshold) voltages,
$V_L$ and $V_R$ are threshold voltages, and  $M_1$ and $M_2$ are limiting
values of the memristance $M$. In Eq. (\ref{eq2}), the $\theta$-functions
guarantees that the memristance changes only in the interval between
$M_1$ and $M_2$. The shape of $f(V)$ is sketched in \figref{Equivalence}.

The only difference between the memristive device described by Eqs. (\ref{eq1})-(\ref{eq3}) and the one used in our previous work \cite{Amoeba_yuriy} is the presence of asymmetric voltage thresholds (the previous model \cite{Amoeba_yuriy}  is obtained by setting $V_L=V_R$ in Eqs. (\ref{eq1})-(\ref{eq3})). Such a modification is needed in order to avoid
 unwanted circuit dynamics. In particular, if a long train of voltage pulses with a resonance frequency was applied to the LC-contour (shown in \insref{Circuits}{a}) with a symmetric memristive device, the amplitude of oscillations across the capacitor would increase driving a sequence of unwanted switchings between $M_1$ to $M_2$. However, these switchings (interpreted as fast learning and unlearning processes) are highly unlikely to be observed in {\it Physarum polycephalum}.

In electronics, there are at least two ways to realize a memristive device with asymmetric switching threshold. One approach is to select an experimental memristive system with required characteristics from a large amount of presently known memristive systems \cite{pershin11a}. Another approach consists in the asymmetrization of a symmetric memristive system via its coupling to a non-linear circuit element. For example, it can be performed by attaching a resistor in parallel to a diode to a terminal of a memristive system with symmetric thresholds. The resulting three elements circuit can be considered as a single effective memristive device with switching asymmetry (see Fig.~\ref{Equivalence}).

In fact, many emergent non-volatile memory cells \cite{waser07a,scott07a,sawa08a,karg08a,burr08a,Yang_2008,pershin11a} seems perfect for analog applications in electronics (including those shown in Fig. \ref{Circuits}) because of their long-term information storage capability, threshold-type switching, high endurance, low power consumption and short read/write times (normally, all these characteristics are desirable for non-traditional computing applications \cite{pershin12a}).  Some of these systems, e.g., nanoionic resistive switches based on amorphous-Si (see Ref. \cite{jo09a}), offer an additional benefit: a CMOS compatibility. Importantly, examples of memristive devices with asymmetric switching thresholds are known, such as Cu/SiO$_2$/Pt electrochemical metallization cells \cite{Schindler09b}.

\begin{figure}
\centerline{
\includegraphics[width=0.9\columnwidth]{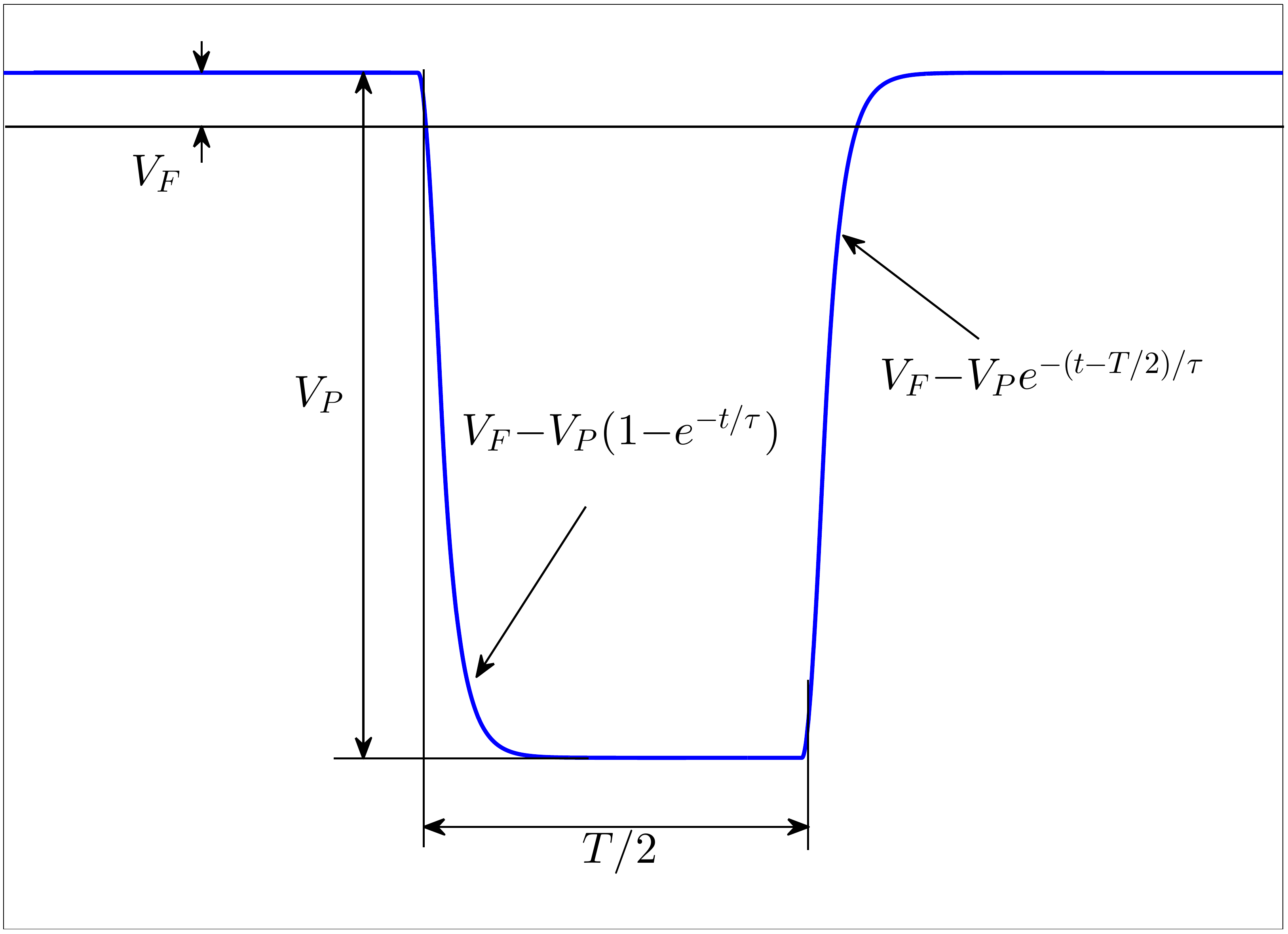}}
\caption{\label{impulse}The shape of the voltage pulse describing a pulse of unfavorable environmental conditions.}
\end{figure}

In our circuit simulations, the environmental conditions are described by a  single parameter, the applied voltage $V(t)$. While the standard (favorable) environmental condition corresponds to a slightly positive $V(t)$, namely $V(t)=V_F$, the unfavorable conditions are simulated by a negative applied voltage. The choice for the shape of unfavorable conditions pulse (a short term variation of temperature and humidity) is based on a model of an object in heat contact with variable temperature reservoir, which is indeed what is most likely to be done in actual
biological experiments. The pulse shape parameters are given in \figref{impulse}.

The circuits from \figref{Circuits} respond differently to favorable and unfavorable conditions.
In the case of favorable conditions applied for a long period of time, memristive elements are subjected to a small positive voltage that, according to Eqs. (\ref{eq1})-(\ref{eq3}), switches these elements into the low resistance state $M_1$. In this case, oscillations in LC contour(s) are damped. On the other hand, pulses of unfavorable conditions induce a complex circuit dynamics depending on many factors such as the amplitude of the pulse $V_P$, the pulse length $T/2$, separation between the pulses, etc. Ref. \cite{Amoeba_yuriy} analyzes the response of \insref{Circuits}{a} circuit to a periodic and aperiodic sequences of three pulses. It is shown that only the periodic sequence with a frequency close to the fundamental frequency of LC contour drives the circuit into the undamped state. Such a behavior is related to the fact that when resonant pulses are applied, the amplitude of voltage oscillations on the capacitor increases with each pulse and at some point exceeds the threshold voltage of the memristive device that switches into the high resistance state.

Finally, we mention that another way to include memory loss mechanisms independently of the environmental condition, can be done by employing a memristive system with an internal memory decay. In order to take into account the internal memory decay, \Eqref{eq3} can be replaced, for example, with
\begin{numcases} {f(V)=}
\beta\left(V-V_R\right) -\gamma x & for $V\geq V_R$  \nonumber
\\
\beta\left(V+V_L\right) -\gamma x & for $V\leq -V_L$  \;\;\; \;\;\;\;\; .\label{Vcontr2}
\\
-\gamma x & otherwise  \nonumber
\end{numcases}

\begin{figure*}
\centerline{
\includegraphics[width=2\columnwidth]{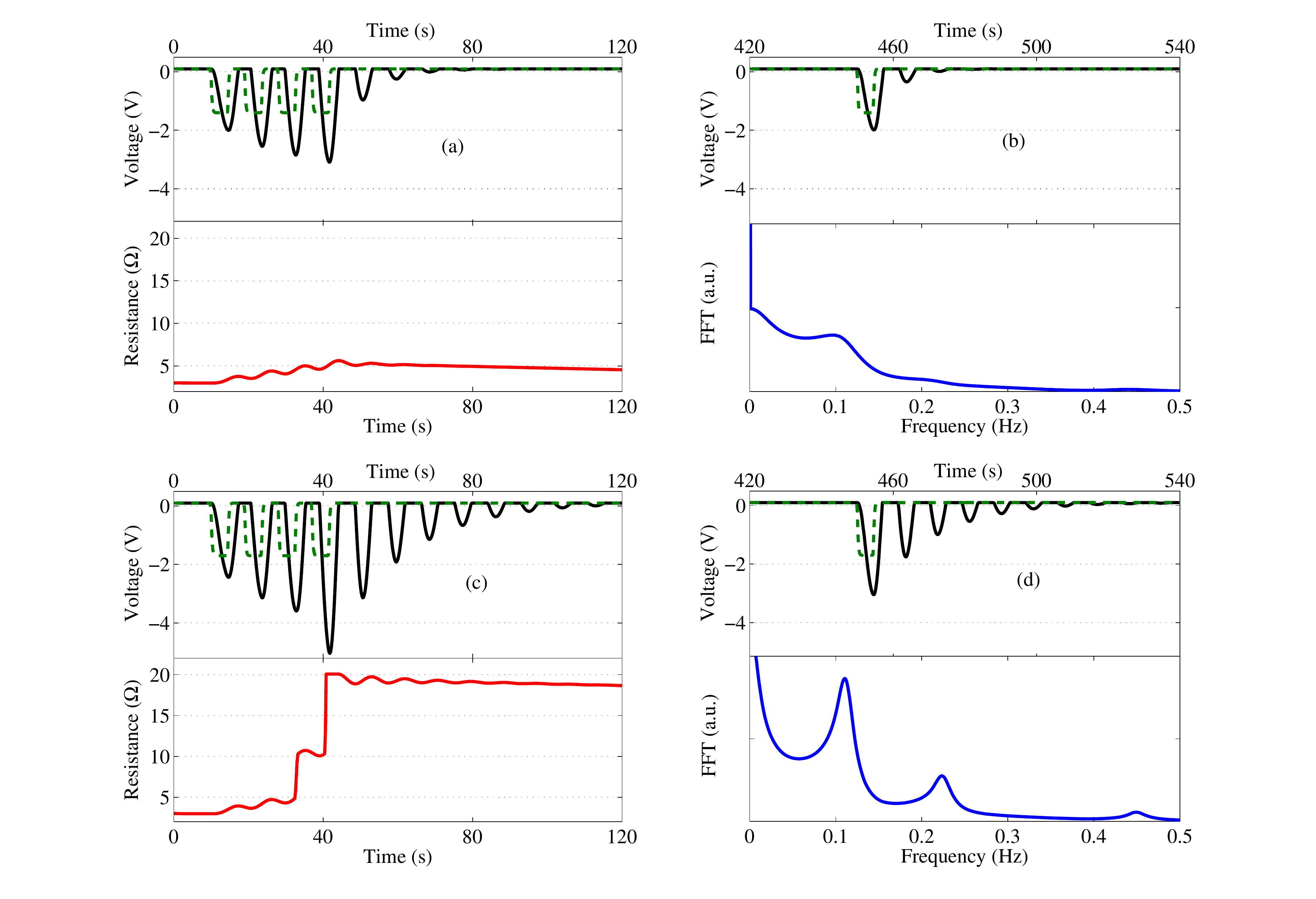}}
\caption{\label{response_4imp} Simulation of the single LC contour (shown \insref{Circuits}{a}) subjected to a periodic sequence of pulses and tested by a single pulse at a longer time. These plots have been obtained with $R=0.1\Omega$, $L=2\textnormal{H}$, $C=1\textnormal{F}$, $M_1=3\Omega$, $M_2=20\Omega$, $\alpha=0.1\Omega(\textnormal{Vs})$, $\beta=100\Omega/(\textnormal{Vs})$, $V_L=3.5\textnormal{V}$ and $V_R=10.5\textnormal{V}$. The voltage corresponding to the favorable condition was $V_F=0.1\textnormal{V}$. (a) and (b) show the response to below-the-threshold pulse sequence, $T=9\textnormal{s}$ and $V_P=1.6\textnormal{V}$. c) and d) show response to above-the-threshold (learning) pulse sequence response, $T=9\textnormal{s}$ and $V_P=1.8\textnormal{V}$. The input signal is shown by dashed green lines.}
\end{figure*}

Here, $\gamma$ is the relaxation constant responsible for the memory decay, the meaning of all other parameters is the same. It is worth noticing that memristive systems with internal relaxation are well known. For example, spin memristive systems \cite{pershin08a} have a short-term memory. In our case study the favorable condition does not vary, so \Eqref{eq3} and \Eqref{Vcontr2} give essentially the same results for a suitable choice of $\gamma$.

Using the memristive device model with threshold asymmetry given by Eqs. (\ref{eq1})-(\ref{eq3}), we have performed simulations of circuits presented in \figref{Circuits}. We use the in-house NOSTOS (NOnlinear circuit and SysTem Orbit Stability) simulator developed by one of the authors (FT)  \cite{TraversaIJCTA,TraversaAEU,TraversaTCAD,Traversa_IET}. A qualitatively similar response of all circuits to periodic and aperiodic sequences of three pulses (corresponding to the protocol of the biological experiment \cite{Saigusa_2008}) has been found providing that, initially, the internal frequency of the adaptive contour is close to the pulse period (see  Sec. \ref{sec4} for more details). Applying different (more complex) training sequences, we have been able to determine cases when responses become circuit-specific. We will focus below on such cases including the excitations by multiple frequency sequences and tests of learning threshold.

%%%%%%%%%%%%%%%%%%%%%%%%%%%%%%%%%%%%%%%%%%%%%%%%%%%%%%%%%%%%%
%%%%%%%%%%%%%%%%%%%%%%%%%%%%%%%%%%%%%%%%%%%%%%%%%%%%%%%%%%%%%
\section{Coupled LC-contours with memristive damping} \label{sec3}
%%%%%%%%%%%%%%%%%%%%%%%%%%%%%%%%%%%%%%%%%%%%%%%%%%%%%%%%%%%%%
%%%%%%%%%%%%%%%%%%%%%%%%%%%%%%%%%%%%%%%%%%%%%%%%%%%%%%%%%%%%%

In this Section, we consider the response of the circuit presented in \insref{Circuits}{b} to different pulse sequences.
The choice of pulse sequences is dictated by our desire to better understand/predict the slime mold's response and know
implications of different models.

%%%%%%%%%%%%%%%%%%%%%%%%%%%%%%%%%%%%%%%%%%%%%%%%%%%%%%%%%%%%%
%%%%%%%%%%%%%%%%%%%%%%%%%%%%%%%%%%%%%%%%%%%%%%%%%%%%%%%%%%%%%
\subsection{Testing the learning threshold} \label{sec:tr}
%%%%%%%%%%%%%%%%%%%%%%%%%%%%%%%%%%%%%%%%%%%%%%%%%%%%%%%%%%%%%
%%%%%%%%%%%%%%%%%%%%%%%%%%%%%%%%%%%%%%%%%%%%%%%%%%%%%%%%%%%%%

\begin{figure}
\centerline{
\includegraphics[width=1\columnwidth]{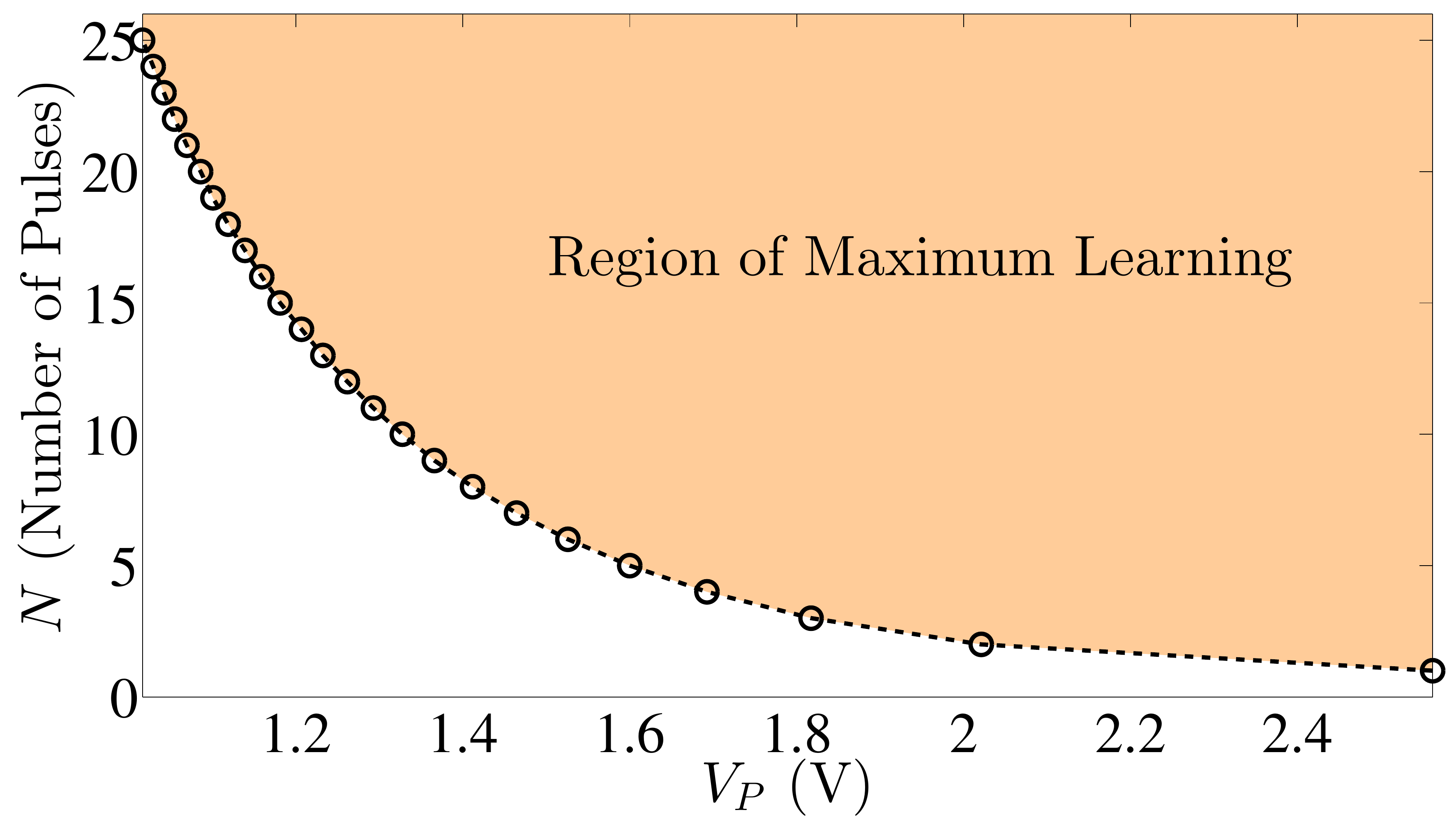}}
\caption{\label{NP_vs_Vp}Learning region of LC contour. The parameters used in the simulation are the same as in \figref{response_4imp}.}
\end{figure}

According to {\bf S1}, there is a threshold for learning and, as a consequence, relatively small changes in environmental conditions can not be memorized. This assumption is based on the fact that virtually any information storage is  associated with a threshold. Moreover, biochemical/physical processes in biological organisms leading to memory (such as chemical reactions) often involve a threshold barrier to overcome. In order to test this assumption, we suggest to study the amoeba's response utilizing a fixed frequency sequence at several pulse amplitudes and number of pulses.

We have found that when a fixed frequency pulse sequence is applied, the response of all circuits in \figref{Circuits} is qualitatively similar providing that, initially, the internal frequency of the adaptive contour of \insref{Circuits}{c} is close to the pulse period. Moreover, we assume that for the circuit of \insref{Circuits}{b} the distribution of the internal resonance frequencies is dense in a given window of frequencies. This avoids different responses that may appear if the pulse frequency is at the boundary between internal frequencies of two contours because we assume such behavior is inconsistent with the amoeba dynamic response. Therefore, we report only simulations for the simplest circuit -- the single LC contour with memristive damping shown in \insref{Circuits}{a} -- assuming that the pulse frequency is close to its internal frequency. To better relate to the amoeba experiments, when plotting the response of the circuit, we impose a restriction so that the response signal (the voltage on the capacitor C in \insref{Circuits}{a}) cannot exceed a certain value, which in our particular calculations is selected to be equal to the voltage corresponding to standard favorable conditions $V_F$ (see Ref. \cite{Amoeba_yuriy} for more details). Electronically, such a response can be obtained using a diode connected from one side to the junction of the inductor, capacitor, and memristive system in \insref{Circuits}{a} and a (large) resistor connected between another side of the diode and a power source at $V_F$ voltage.

\figref{response_4imp} reports the response of the \insref{Circuits}{a} circuit trained by a sequence of four pulses. The pulse period ($T=9\textnormal{s}$) is chosen close to the internal frequency of the contour $f_r\approx(2\pi\sqrt{CL})^{-1}$. In particular,  \insref{response_4imp}{a} and (b) indicate that smaller amplitude pulses ($V_P=1.6\textnormal{V}$) do not produce significant switching (see the bottom panel of \insref{response_4imp}{a}) keeping the oscillations strongly damped. This is reasonable since when the amplitude of learning pulses is small, the voltage across the memristive device does not exceed the threshold voltage $V_L=3.5\textnormal{V}$ and the memristance $M$ is only slightly increased towards $M_2$. Damped oscillations of the response signal induced by a testing pulse at a longer time (\insref{response_4imp}{b} panel) are clearly reflected in the Fourier transform of the output signal (see the bottom panel of (b)) that does not contain any sharp peaks.

By contrast, at higher pulse amplitudes, the circuit response becomes similar to the one observed in biological experiments \cite{Saigusa_2008}. In particular,
a spontaneous in-phase slowdown (SPS) \cite{Saigusa_2008} and an SPS after one disappearance event \cite{Saigusa_2008} can be clearly recognized in \insref{response_4imp}{c} and (d), respectively. Such a change in response is a consequence of the fact that now the voltage across the memristive device exceeds the threshold voltage $V_L$ during the training phase (\insref{response_4imp}{c}). In this case, the memristive device is brought into the high resistance state $M=M_2$ (see the bottom panel of \insref{response_4imp}{c}). FFT of the response signal induced by a testing pulse at a longer time (\insref{response_4imp}{d}) exhibit a narrow peak close to $f_r$ and its multiples.

The number of pulses $N$ in the training sequence is an additional parameter controlling the learning. We define the \emph{Region of Maximum Learning} (RML) as the region in the $V_P-N$ plane where the switching of the memristive device is maximum (the condition $M=M_2$ takes place during the training phase). It should be emphasized that the learning process is a continuum analog process, so that the memory build-up is gradual. \figref{NP_vs_Vp} depicts RML for a single LC contour. It is worth noticing that, in principle, such a region can be easily found experimentally for real biological organisms like {\it Physarum polycephalum}.  Then, the memristive device model could be tailored to attain a better match with behavior of specific biological organisms.

%%%%%%%%%%%%%%%%%%%%%%%%%%%%%%%%%%%%%%%%%%%%%%%%%%%%%%%%%%%%%
%%%%%%%%%%%%%%%%%%%%%%%%%%%%%%%%%%%%%%%%%%%%%%%%%%%%%%%%%%%%%
\subsection{Multiple frequency response}
%%%%%%%%%%%%%%%%%%%%%%%%%%%%%%%%%%%%%%%%%%%%%%%%%%%%%%%%%%%%%
%%%%%%%%%%%%%%%%%%%%%%%%%%%%%%%%%%%%%%%%%%%%%%%%%%%%%%%%%%%%%

\begin{figure}
\centerline{
\includegraphics[width=1\columnwidth]{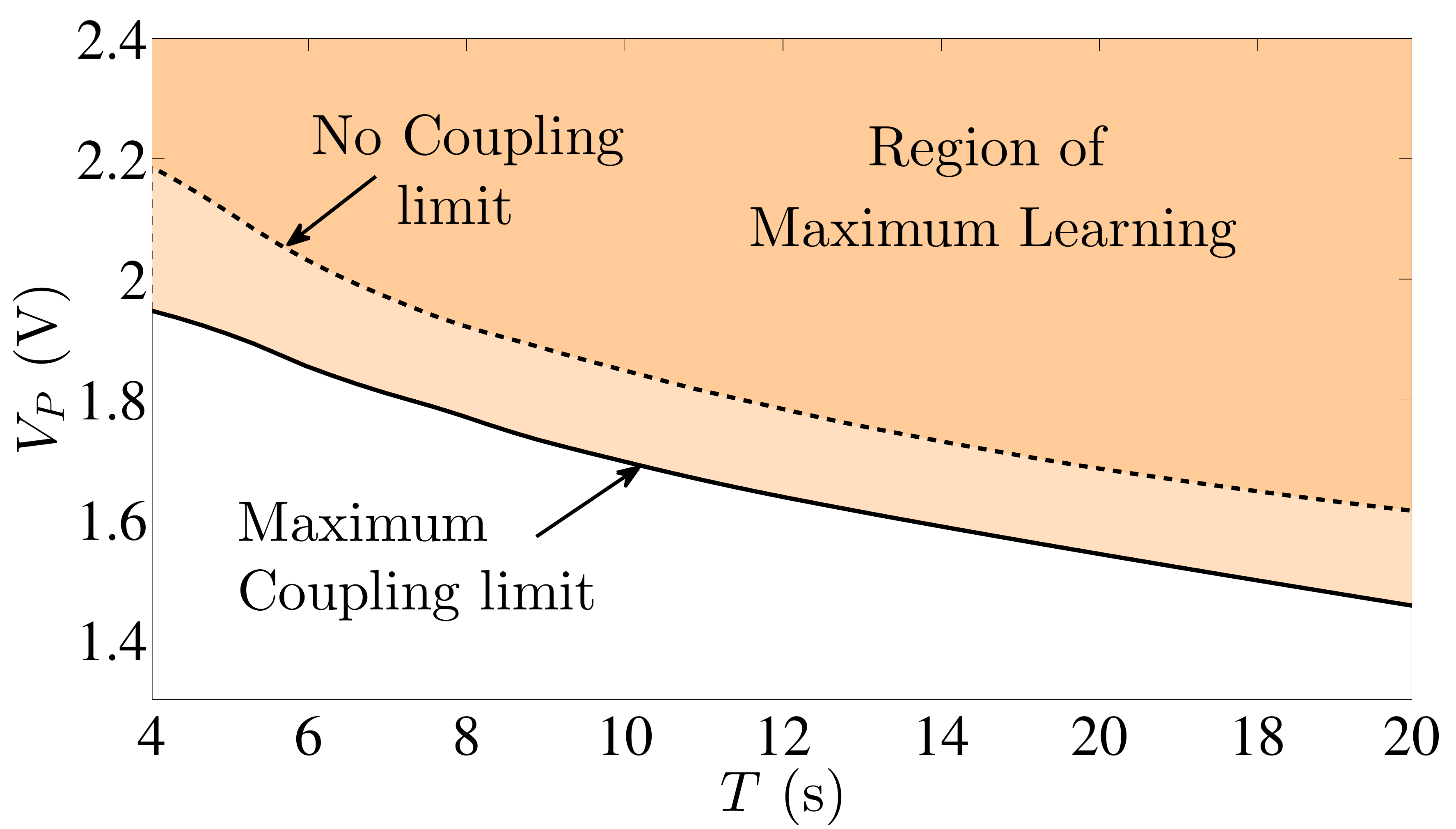}}
\caption{\label{T_vs_Vp}RML for $n$ coupled LC contours (\insref{Circuits}{b}). The dashed line confines the RML for $R_c=0 \hspace{0.02in}\Omega$ and the solid line for $R_c=0.1 \hspace{0.02in}\Omega$. The calculations were made using the following system parameters: $R=0.1\Omega$, $L=2h\textnormal{H}$, $C=h\textnormal{F}$, $M_1=3\Omega$, $M_2=50\Omega$, $\alpha=0.1\Omega/(\textnormal{Vs})$, $\beta=100\Omega/(\textnormal{Vs})$, $V_L=3.5\textnormal{V}$ and $V_R=10.5\textnormal{V}$ where $h$ takes values in the interval $[0.5, 2]$ logarithmically spaced in $12$ points.}
\end{figure}

\begin{figure}
\centerline{
\includegraphics[width=1\columnwidth]{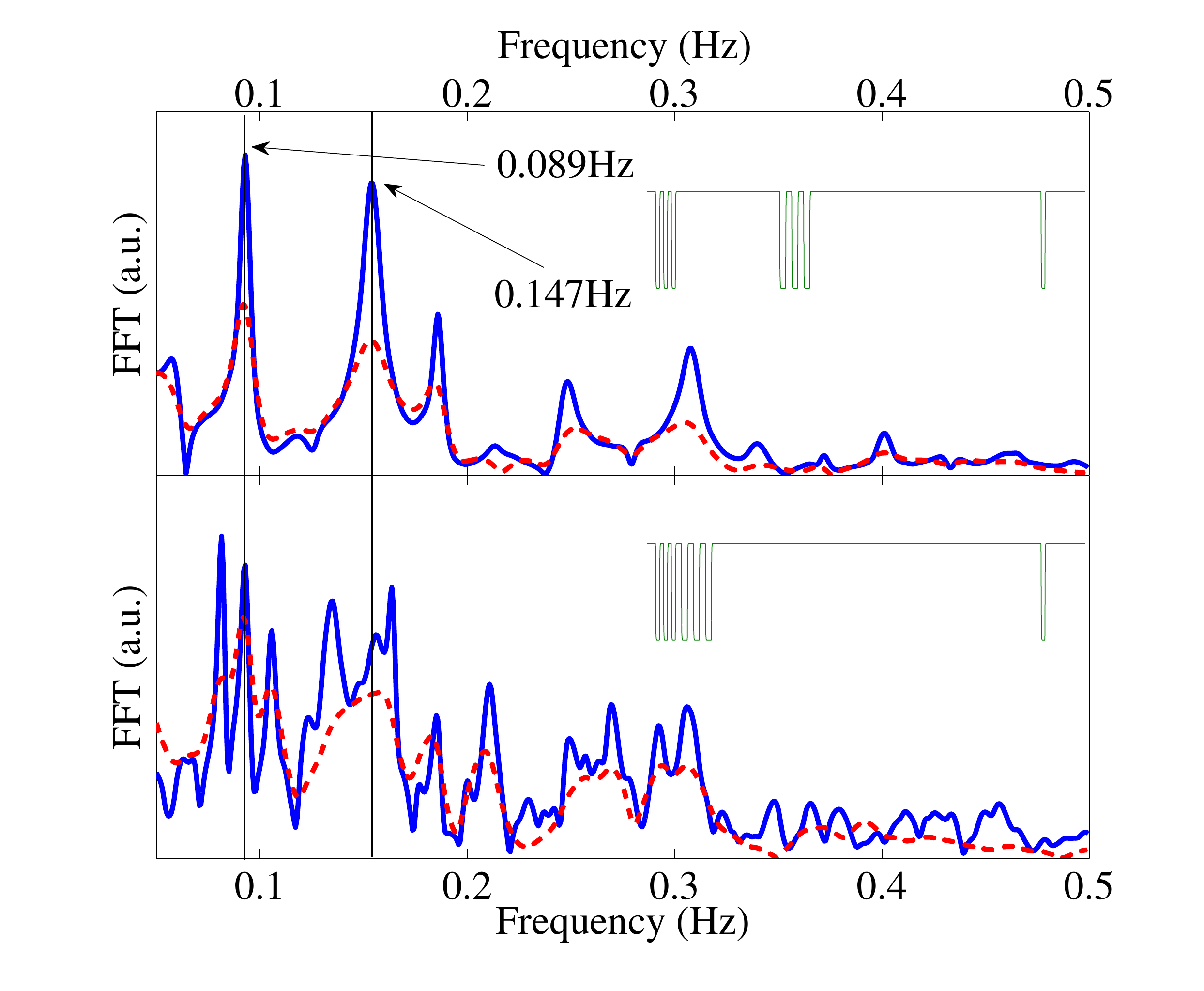}}
\caption{\label{multi_f}Fourier transform of the voltage response after the application of two pulse sequences as described in the text. The calculations use the same parameters as in \figref{T_vs_Vp}. The voltage corresponding to the favorable condition is $V_F=0.1\textnormal{V}$. The periods used for the two trains  are $T_1=6.803\textnormal{s}$ and $T_2=11.261\textnormal{s}$, respectively. The corresponding frequencies are $f_1=0.147\textnormal{Hz}$ and $f_2=0.089\textnormal{s}$. The dashed red lines are the response at $R_c=0\textnormal{V}$ (no coupling) with pulse amplitudes of $V_{P_1}=1.95\textnormal{V}$ and $V_{P_2}=1.7\textnormal{V}$. The solid blue lines correspond to $R_c=0.1\textnormal{V}$ (maximum coupling) with pulse amplitudes of $V_{P_1}=1.85\textnormal{V}$ and $V_{P_2}=1.65\textnormal{V}$. The top panel is obtained with $T_s=150\textnormal{s}$ time separation between pulse sequences, $T_s=0\textnormal{s}$ for the bottom panel. }
\end{figure}

Additional information on internal processes leading to learning and memory in simple biological organisms can be obtained using
more complex pulse sequences combining two or more frequencies. Moreover, the knowledge of biological response to such sequences would help to understand whether \insref{Circuits}{b}  circuit or \insref{Circuits}{c} circuit is a closer
representation of the real dynamics. Specifically, while \insref{Circuits}{c} circuit can learn only a single frequency, \insref{Circuits}{b} circuit can memorize many of these. The frequency memory can be tested using an SPS after one disappearance event experiment \cite{Saigusa_2008}.

Let us now consider a circuit consisting of $n$ coupled LC contours with memristive damping as shown in \insref{Circuits}{b}. The output signal of this circuit (corresponding to the locomotion speed of slime molds) is the minimal of voltages on the capacitors C$_i$ with a restriction that the response cannot exceed a certain value selected as $V_F$. Electronically,
such a response signal can be obtained using $n$ diodes, each connected from one side to the junction of the inductor, capacitor, and memristive device in \insref{Circuits}{b} and, from the second side, to a common connection of all diodes with a (large) resistor connected (using its second terminal) to a power source at $V_F$ voltage.

Using similar considerations as in Sec. \ref{sec:tr}, we plot RML for $n$ coupled LC contours. Fig. \ref{T_vs_Vp} presents this result for two values of $R_c$. Note, that the value of the coupling parameter, $R_c$ (see the circuit in \insref{Circuits}{b}), is responsible for a shift of the boundary of RML.

Based on information regarding the location of the RML region, we have applied a voltage input composed of two sequences of three pulses of different periods and amplitudes. In particular, by fixing two distinct periods, the corresponding amplitudes have been taken corresponding to the coupling limits of the RML in \figref{T_vs_Vp}. The initial separation $T_s$ between two sequences has been selected long enough so that the oscillations due to the first sequence are damped before the start of the second one. The top panel of \figref{multi_f} presents the Fourier transform of the response to a separate testing pulse after the application of training pulses described above. In this case, the spectrum clearly shows the presence of two peaks corresponding to two training frequencies, and also to their sum and difference and to their second harmonics. This spectrum shape is essentially due to the fact that only two memristive systems in the contours with resonant frequencies close to those of the input sequences are switched into $M_2$. Note that peaks are sharper at stronger coupling between the contours.

Reducing the time-separation $T_s$ between two sequences, the response of the circuit dramatically changes. The bottom panel in \figref{multi_f} reports the Fourier transform of the response to a separate testing pulse after the training by two sequences with $T_s=0\textnormal{s}$. Compared to the top panel, the number of peaks has increased. The presence of many more peaks is due to switching of many memristive systems. The reason is that two consecutive pulse sequences induce large enough oscillations, not just in resonant contours.

%%%%%%%%%%%%%%%%%%%%%%%%%%%%%%%%%%%%%%%%%%%%%%%%%%%%%%%%%%%%%
%%%%%%%%%%%%%%%%%%%%%%%%%%%%%%%%%%%%%%%%%%%%%%%%%%%%%%%%%%%%%
\section{Adaptive contour}  \label{sec4}
%%%%%%%%%%%%%%%%%%%%%%%%%%%%%%%%%%%%%%%%%%%%%%%%%%%%%%%%%%%%%
%%%%%%%%%%%%%%%%%%%%%%%%%%%%%%%%%%%%%%%%%%%%%%%%%%%%%%%%%%%%%

This Section discusses possible realizations of the adaptive contour shown in \insref{Circuits}{c}, namely, a single memcapacitive device-based oscillating contour that can adapt its frequency to the frequency of the applied signal.
The general problem can be stated in the following way: let a circuit be described by a dynamical system
\begin{equation}
\frac{dq(x,y)}{dt}=f(x,y)+s(t) \label{dynamical}  \;\; ,
\end{equation}
where $x$ is the vector of circuit state variables, namely, currents and potentials, $y$ is the vector composed of internal states variables of memelements (in our case memristors and memcapacitors)~\cite{diventra09a}, $q$ and $f$ are vector functions describing the circuit equations, and $s$ is the external input. Considering a periodic signal $s$ of frequency $\omega$, we say that the circuit is \emph{frequency adapting} if there is a non-empty subset of elements of $y$ organized in a vector $y_a$ such that
\begin{equation}
\lim_{t\rightarrow\infty}y_a(t) = \bar y_a(\omega)+h(t) \;\; ,  \label{lim}
\end{equation}
where $||h(t)||\ll ||\bar y_a(\omega)||$, i.e., $||y_a||$ reaches a quasi-constant value depending on the external frequency. In addition, for the LC contour of \insref{Circuits}{c}, we also require that $C=C(y_a)$, i.e., the memcapacitance has to reach a quasi-constant value according to the external frequency. Moreover, the dependence $C(y_a)$ must provide a shift of the contour resonant frequency to the external frequency $\omega$.
It is worth noticing that the main difference between functionalities of frequency adaptive contours considered in this section and the coupled contour system of \insref{Circuits}{b} is that the former can adapt to one resonant frequency at a time, while the latter can activate more resonant frequencies at the same time.

Generally, it is difficult to determine a model of memcapacitive device providing such frequency-adaptive functionality. More difficult, however, is to find experimental solid-state realizations of such model. The latter task is, however, beyond the scope of this paper. The difficulties with memcapacitive models stem from the fact that the resulting systems depend on several internal state variables \cite{diventra09a} with dynamics described by a set of nonlinear differential equations such that the dynamics can be unstable or can show some unexpected behavior. Here, we will discuss two different models of memcapacitive devices able to change the capacitance according to the external frequency. Being the models dependent on several parameters, an exhaustive picture of their dynamics would be out of the scope of this paper. For this reason, we only discuss some specific sets of parameters that minimally realize the experimental facts ({\bf F}). Ideally, we would like to formulate a model resulting in a stable dynamics and fast frequency learning.

\subsection{Effective model of certain memcapacitive systems}\label{effect}

Before giving the description and results of the two memcapacitive models employed in this section, we briefly discuss the general equations that govern charge-controlled memcapacitive systems \cite{diventra09a}.  We also discuss under which conditions it is possible to derive the equations of an {\it effective} charge- and current-controlled memcapacitive system starting from a charge-controlled one. This result is especially useful when including charge-controlled memcapacitive system models in commercial simulators that generally use currents and voltages as unknowns.

A charge-controlled memcapacitive system is a device governed by the equations \cite{diventra09a}
\begin{eqnarray}
V &=& C(\mathbf x, q, t)^{-1}q, \label{mem1}\\
\dot{\mathbf x} &=& f(\mathbf x,q,t),\label{mem2}
\end{eqnarray}
where $\mathbf x$ is an $m$-component vector casting all the internal state variables, $q$ is the charge, $V$ is the applied voltage, and $f(\mathbf x,q,t)$ is the $m$-component vector function describing the evolution of internal state variables. We note that certain $m$-order charge-controlled memcapacitive systems can be effectively described as $(m-1)$-order charge- and current-controlled memcapacitive systems (similar transformations can be made for all types of memory circuit elements \cite{diventra09a}). In particular, let us assume that Eq. (\ref{mem2}) for $x_m$ is given by
\begin{equation}
\dot{x}_m=-\frac{x_m-q(t)}{\tau} , \label{eq:xm}
\end{equation}
where $\tau$ is a (small) time constant (the change of $q$ on the time scale of $\tau$ should be small). Then, the solution of Eq. (\ref{eq:xm}) can be written as
\begin{equation}
x_m=\frac{1}{\tau}\int\limits_{-\infty}^0e^{\frac{\tilde{t}}{\tau}}q\left( \tilde{t}+t \right) \textnormal{d}\tilde{t}+C_1e^{-\frac{t}{\tau}}. \label{eq:xm1}
\end{equation}
Here, $C_1$ is the integration constant that is set equal to zero. Because of the exponential term under the integral in Eq. (\ref{eq:xm1}) and the smallness of $\tau$, the main contribution to the integral is provided at $\tilde{t}$ close to zero. We then Taylor expand $q\left( \tilde{t}+t \right)$ around $\tilde{t}=0$ and integrate over $\tilde{t}$ to obtain
\begin{equation}
x_m=q(t)-\tau\frac{\textnormal{d}q}{\textnormal{d}t}+\tau^2 \frac{\textnormal{d}^2q}{\textnormal{d}t^2}-\tau^3 \frac{\textnormal{d}^3q}{\textnormal{d}t^3}+ .... \label{eq:xm2}
\end{equation}

Clearly, in the limit of small $\tau$ and well-behaving $q(t)$, the combination $\left(q(t)-x_m\right)/\tau$ is the current $i\equiv \textnormal{d}q / \textnormal{d}t$. Thus (see Eqs. (\ref{mem1}) and (\ref{mem2})), both $C$ and $f$ can be considered as functions of the current $i$. This fact can be written explicitly omitting $x_m$ from the set of internal state variables. Consequently, Eqs. (\ref{mem1}), (\ref{mem2}) can be rewritten in the form of an effective $(m-1)$-order charge- and current-controlled memcapacitive system
\begin{eqnarray}
V &=& C_{eff}( \tilde{\mathbf x}, q, i, t)^{-1}q, \label{mem1a}\\
\dot{\tilde{\mathbf x}} &=& f_{eff}(\tilde{\mathbf x}, q, i,t),\label{mem2a}
\end{eqnarray}
where $\tilde{\mathbf x}$ is the set of $m-1$ internal state variables, $C_{eff}( \tilde{\mathbf x}, q, i, t)$ is the effective memcapacitance and $f_{eff}(\mathbf x,q,i,t)$ is the effective $(m-1)$-component vector function describing the evolution of internal state variables. We found that effective current-controlled memcapacitive models are quite useful in studies of adaptive frequency functionality. The reduced description of memcapacitive systems based on Eqs. (\ref{mem1a}) and (\ref{mem2a}) is adopted in the two models discussed below.

%We now assume that there is a linear function $g$ such that
%\begin{equation}
%\mathbf y=g(\mathbf x,q)=K\mathbf x+q\mathbf 1,
%\end{equation}
%where $\mathbf y$ is of the same size of $\mathbf x$, $K$ a $m\times m$ matrix and $\mathbf 1$ a $m-$component vector with all unity entries and that the following conditions are satisfied:
%\begin{description}
%\item[a)] $K$ is invertible, i.e. the linear function $h$ exists such that $\mathbf x = h(\mathbf y,q)=K^{-1}(\mathbf y-q\mathbf 1)$,
%\item[b)] $\frac{\textnormal{d}}{\textnormal{d}t} [q/C(h(\mathbf y,q), q, t)]=i/C_{eff}(\dot{\mathbf y},\mathbf y, i, t) $ where $i=\dot q$,
%\item[c)] $f(\mathbf x,q,t)=f(g(\mathbf x,q),t)=f(\mathbf y,t)$.
%\end{description}
%When the above mentioned conditions are satisfied, \Eqref{mem1} and \Eqref{mem2} can be rewritten as
%\begin{eqnarray}
%C_{eff}(\dot{\mathbf y},\mathbf y, i, t) \dot V &=& i, \label{mem1_bis}\\
%\dot{\mathbf y} =i\mathbf 1 + K f(\mathbf y,t) &=& f\rq{}(\mathbf y,i,t) ,\label{mem2_bis}
%\end{eqnarray}

%That can be reinterpreted as an effective current-controlled memcapacitive system. An example of memcapacitive system that satisfies all these condition is the super-lattice memcapacitive system discussed in Ref. \cite{Martinez}. Finally, it is worth noticing that by making one more invertible variable change $\mathbf z=c(\mathbf y,i)$ in \Eqref{mem1_bis} and \Eqref{mem2_bis} a much more general functions depending on $i$ and $\mathbf z$ instead of the \Eqref{mem2_bis} can be obtained.

%%%%%%%%%%%%%%%%%%%%%%%%%%%%%%%%%%%%%%%%%%%%%%%%%%%%%%%%%%%%%
\subsection{Hopf oscillator memcapacitive system}\label{Hopf}
%%%%%%%%%%%%%%%%%%%%%%%%%%%%%%%%%%%%%%%%%%%%%%%%%%%%%%%%%%%%%

\begin{figure}
\centerline{
\includegraphics[width=1\columnwidth]{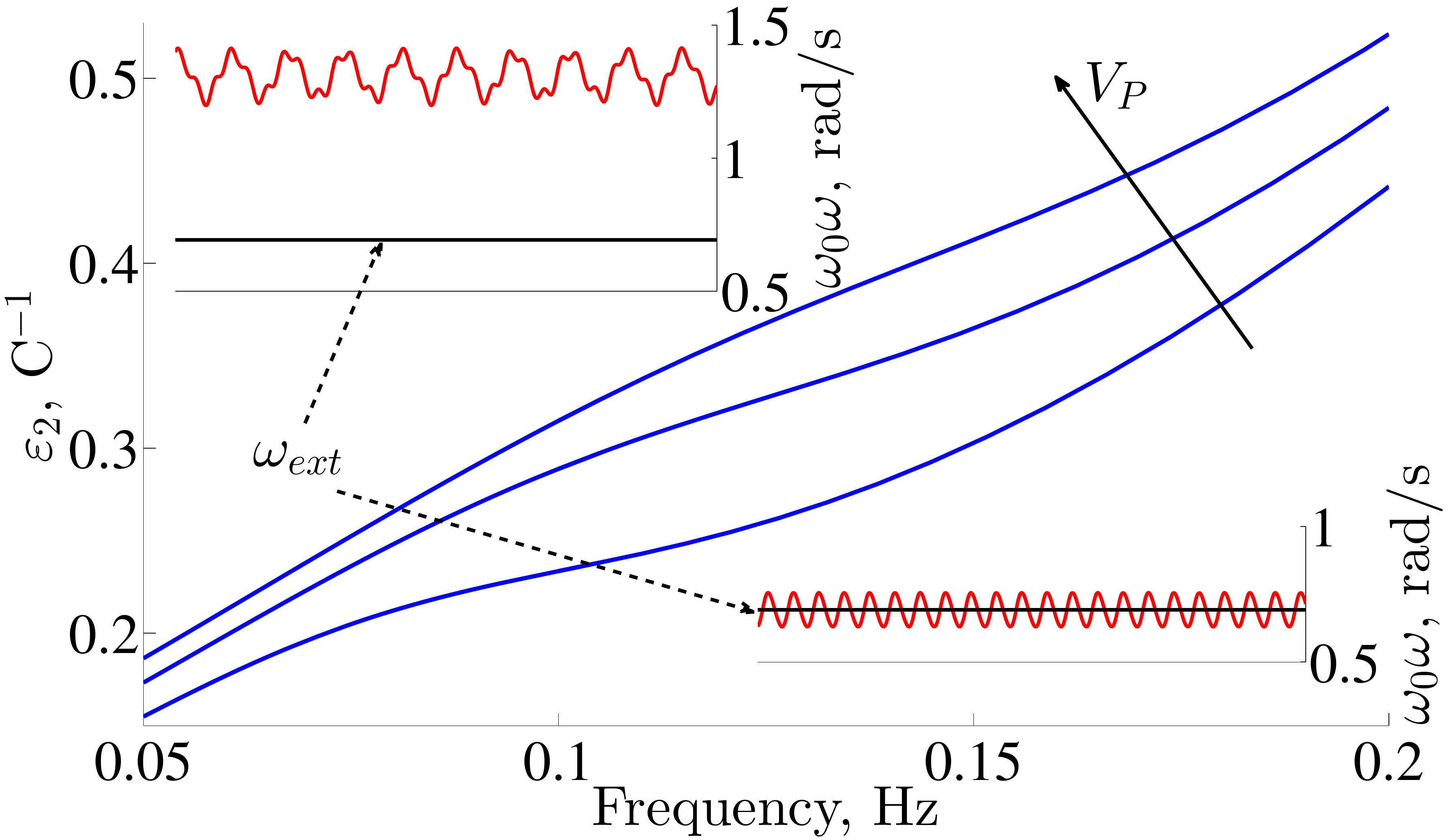}}
\caption{\label{Hopf_bif}Bifurcation curves of the circuit of \insref{Circuits}{c} obtained with Hopf oscillator model of memcapacitive system. The circuit and memristive system  parameters are the same as in \figref{T_vs_Vp}. The parameters of memcapacitive system are: $\gamma=10 \hspace{0.01in}\textnormal{s}^{-1}$, $\varepsilon_1=0.5 \hspace{0.01in}\textnormal{C}^{-1}$, $\omega_0=0.69 \hspace{0.01in}\textnormal{rad}/\textnormal{s}$, $\alpha=10^{-3}$ and $\mu=1$. The bifurcation curves were tracked at $V_P=1,2,3\textnormal{V}$. The bottom and top insets show the stable steady states of the variable $\omega$  before and after the bifurcation, respectively, for an external frequency $\omega_{ext}=\omega_0$ and $\varepsilon_2=0.05 \textnormal{C}^{-1}$ and $0.35 \hspace{0.01in} \textnormal{C}^{-1}$.}
\end{figure}
\begin{figure}
\centerline{
\includegraphics[width=1\columnwidth]{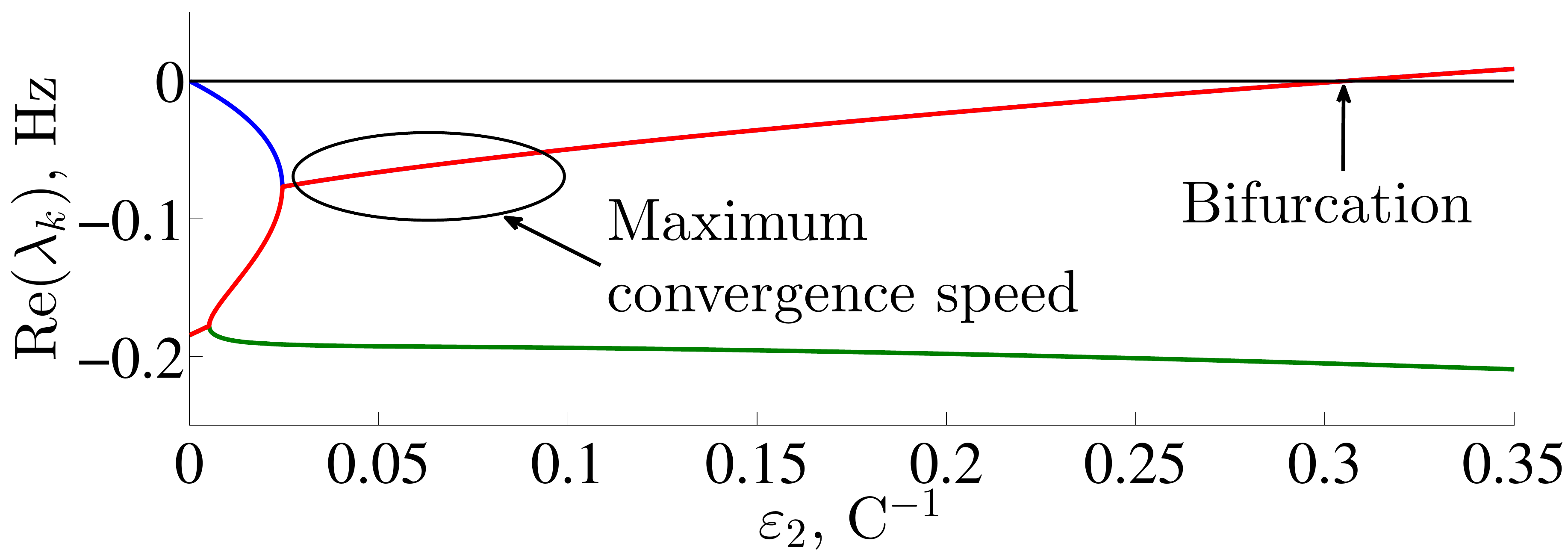}}
\caption{\label{Hopf_Floquet}Three Floquet exponents with the largest real part for $V_P=1\textnormal{V}$ and $\omega_{ext}=0.69\textnormal{rad}/\textnormal{s}$.}
\end{figure}
\begin{figure*}
\centerline{
\includegraphics[width=2\columnwidth]{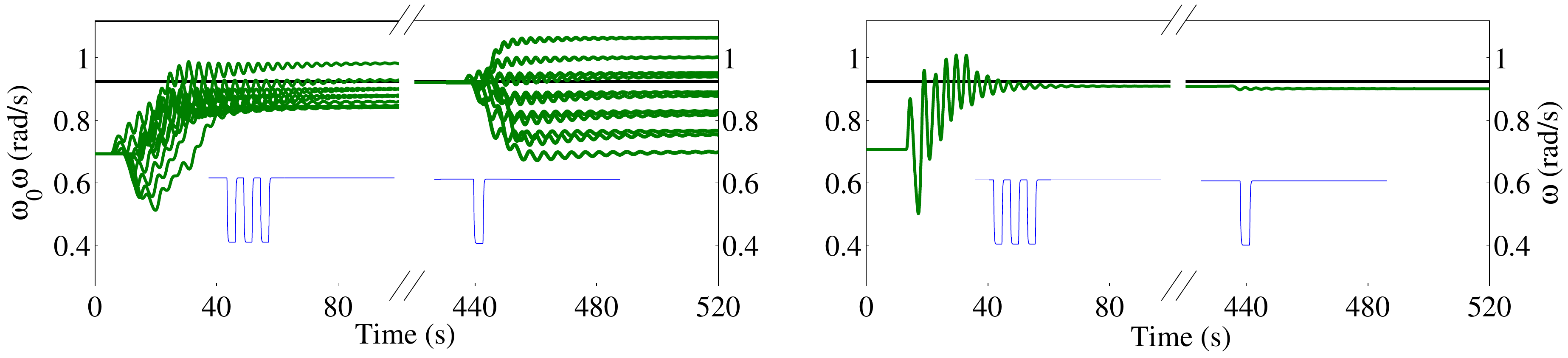}}
\caption{\label{Hopf_PDV_comparison}Simulations of the LC contour from \insref{Circuits}{c} based on Hopf oscillator (left) and DPT (right)
models of memcapacitive system. The left plot was found employing the same parameters as in \figref{Hopf_bif} and $\varepsilon_2=0.05\hspace{0.01in} \textnormal{C}^{-1}$. Different curves correspond to different initial times for the pulse train.
The right plot was obtained utilizing DPT model parameters $\alpha_1=\alpha_2=\alpha_3=10^{-3}\hspace{0.01in} \textnormal{s}^{-1}$, $\beta_1=1\hspace{0.01in} \textnormal{A}^{-1}\textnormal{C}^{-1}$, $\beta_2=1\hspace{0.01in} \textnormal{C}^{-1}$, $\beta_3=1\hspace{0.01in} \textnormal{s}^{-1}$, $c_1=10^{-1}\hspace{0.01in} \textnormal{A}^{-1}\textnormal{C}^{-1}$, and $c_2=10^{-1}\hspace{0.01in} \textnormal{F}^{-1}\textnormal{A}^{-1}\textnormal{C}^{-1}$. The applied voltage was a three-pulse train plus an isolated pulse. In both cases, the pulse parameters are $T=6.8\textnormal{s}$ and $V_P=2.7\textnormal{V}$. The thick black line indicates the external frequency.}
\end{figure*}
The first model of adaptive contour utilizes a memcapacitive system with an internal dynamics similar to that of the Hopf oscillator \cite{Buchli04a,Righetti06a} (analogous to Coram's oscillator in the field of electronics \cite{TraversaTCAS,TraversaFNL}).
We employ the effective (reduced)  formalism of  \Eqref{mem1a} and \Eqref{mem2a} to specify the memcapacitive device.
Formally, such a device is an effective 3-rd order current-controlled memcapacitive system described by the following equations
\begin{eqnarray}
C_{eff}  &=& \frac{1}{L\omega_0^2\omega^2}     \label{Hopf1}\\
\dot x  &=& \gamma(\mu^2-r^2)x-\omega_0\omega y+\varepsilon_1 i     \label{Hopf2}\\
\dot y  &=& \gamma(\mu^2-r^2)x-\omega_0\omega y    \label{Hopf3}\\
\dot\omega  &=& -\varepsilon_2 i\frac{y}{r}-\alpha\omega_0(\omega-1)    \label{Hopf4}
\end{eqnarray}
where $x$, $y$ and $\omega$ are dimensionless internal state variables, $r^2=x^2+y^2$, $i$ is the current, $L$ the inductance of the \emph{LC} contour and $\mu$ and $\alpha$ are dimensionless parameters and $\gamma$, $\varepsilon_1$, $\varepsilon_2$ and $\omega_0$ parameters with proper dimensions. Note, that the internal angular frequency of oscillations in the above model is $\omega_0\omega$. Moreover, we emphasize that in the above equations the current $i$ is more preferable to use compared to the voltage since the DC component of the voltage -- always present in our case study -- induces strong instabilities and/or unbounded increase of the internal variables.

A detailed stability analysis of the Hopf oscillator described by Eqs. (\ref{Hopf2})-(\ref{Hopf4}) at $\omega_0=1\hspace{0.01in}\textnormal{rad}/\textnormal{s}$, $\varepsilon_1=\varepsilon_2$, $\alpha=0$ in the presence of a general external perturbation $i$ is reported in Refs. \cite{Buchli04a,Righetti06a}.
In these publications,  a detailed picture of the system dynamics is obtained by assuming independence of the perturbation term on the system variables $x$, $y$ and $\omega$. It was found that the perturbed system is asymptotically stable, the stationary solution converges to an attractor (which, in the case of a periodic perturbation, is a limit cycle with the same frequency as that of the perturbation) and $\omega_0\omega$ converges to the (dominant) angular frequency of the perturbation. In this situation, $\gamma$, $\varepsilon_1$ and $\varepsilon_2$ determine the speed of convergence to the attractor, $\mu$ is the radius of the unperturbed system, $\alpha$ is a damping constant that, in the absence of perturbation, brings the Hopf oscillator back to the unperturbed limit cycle with radius $\mu$ and frequency $\omega_0$.

Unfortunately, the dynamics of our system can be very different from the dynamics described above because \Eqref{Hopf1} builds a dependence between $x$, $y$, $\omega$ and the current $i$ through other circuit variables. Since an analytical study is very complex in this case, we used a numerical approach based on the Floquet theory to assess the stability of the system and determine the convergence of $\omega_0 \omega$ to the frequency of the applied voltage. The basic questions are whether the limit cycle obtained by applying periodic voltage pulses is an attractor (i.e., the limit cycle is asymptotically stable), and how fast trajectories converge to the attractor. The answer to the second question can help to find model parameters optimizing the speed of the frequency adaptation.

The Floquet exponents of the limit cycle \cite{TraversaIJCTA} contain all necessary information to answer both questions from the previous paragraph. Briefly, by linearizing the system around a limit cycle we obtain a linear differential system with time-periodic coefficients. From Floquet theory, the solution of the linear system is of the form $P(t)\exp[Ft]$, where $P(t)$ a time-periodic matrix and $F$ a constant matrix. The eigenvalues $\lambda_k$ of $F$ are the so-called Floquet exponents. $\textnormal{Re}(\lambda_k)$ determines the stability of the limit cycle: all $\textnormal{Re}(\lambda_k)<0$ lead to asymptotic stability; at least one $\textnormal{Re}(\lambda_k)>0$ leads to instability. Moreover, $\textnormal{Im}(\lambda_k)$ is used to classify bifurcations ($\textnormal{Re}(\lambda_k)=0$):  $\textnormal{Im}(\lambda_k)=0$ is a fold bifurcation, $\textnormal{Im}(\lambda_k)=\pi$ is a flip (period doubling) bifurcation, $\textnormal{Im}(\lambda_k)\neq 0,\pi$ a Neimark-Sacker bifurcation \cite{TraversaIJCTA}. Finally, by definition of Floquet exponents, the smaller $\textnormal{Re}(\lambda_k)$, the faster trajectories converge to the limit cycle.

To calculate the Floquet exponents we used the numerical methods reported in Refs. \cite{TraversaIJCTA,TraversaAEU,TraversaTCAD,Traversa_IET} and included in our NOSTOS simulator. This paper reports only our most important findings because of the large number of model parameters and their combinations. In particular, we found that the exponents do not significantly vary with i) the memristance $M$ and the radius $\mu$, ii) for $\alpha<10^{-2}$, and iii) for $\gamma/\varepsilon_1>5 \hspace{0.01in} \textnormal{Cs}^{-1}$. \figref{Hopf_bif} depicts the bifurcation curves (i.e., the loci where $\textnormal{Re}(\lambda_k)=0$) of the LC contour. The stability of the limit cycle strongly depends on the frequency and the amplitude of the applied voltage and on $\varepsilon_2$. On the one hand, in the stability region (region below the bifurcation curve)  $\omega_0\omega$ approaches the frequency of the voltage source (see bottom inset of \figref{Hopf_bif}) and the attractor is a stable orbit with the same period of the external excitation. On the other hand, in the unstable region (above the bifurcation curve) the variable $\omega$ approaches double the excitation frequency (see top inset of \figref{Hopf_bif}), and the attractor is a stable torus. The reason is that the bifurcation is a Neimark-Sacker bifurcation, and the torus is spanned by a quasi-periodic trajectory with frequencies $n\omega_{ext}+\textnormal{Im}(\lambda_k)$ with $n\in\mathbb{Z}$.

\figref{Hopf_Floquet} shows the first three Floquet exponents with the largest real part. Two of them are completely superposed (complex conjugate exponents), cross 0, and give rise to the bifurcation described above. It is worth noticing that the Floquet exponents have a minimum corresponding to the beginning of the region of maximum convergence speed, as highlighted in the figure. This is a remarkable situation. In fact, \Eqref{Hopf4} suggests that the larger $\varepsilon_2$, the faster the trajectories converge to the limit cycle. On the contrary, the Floquet analysis shows that this is valid only before the minimum. After this minimum the situation is inverted. Finally, we recover the region of maximum convergence speed enveloping all the minima for the frequencies and amplitudes of interest for our case study.

Using the previous stability analysis we are now ready to study the behavior of the circuit from \insref{Circuits}{c} based on the Hopf oscillator memcapacitive system. For this purpose, we utilize a set of parameters for which the circuit is stable and the memcapacitive system is within the maximum convergence speed region. The left inset in \figref{Hopf_PDV_comparison} reports the circuit internal frequency, $\omega_0\omega$, when the circuit is subjected to a three-pulse train plus an isolated pulse sequence. Clearly, the internal frequency $\omega_0\omega$ converges to the external frequency (black solid line) very fast -- with only three pulses. It can be noted from the figure that, depending on the starting instant of the pulse train, different curves originate. This is due to the fact that, depending on the time instant in which the pulse train is applied, the internal variable $y$ takes a value in the interval $[-\mu,\mu]$, so that in \Eqref{Hopf4} the term depending on $i$ can be initially positive or negative, giving rise to the different behaviors. The spread of different curves depends on the number of pulses (its width goes to zero as the number of pulses goes to infinity). Furthermore, as a consequence of the fast convergence, the response of $\omega_0\omega$ is very sensitive to the isolated pulse, as shown in \figref{Hopf_PDV_comparison}. This is also due to the fact that, depending on the time instant in which the pulse starts, the internal variable $y$ takes a value in the interval $[-\mu,\mu]$. This behavior is amplified when the memristive system is switched off because more current can pass through the memcapacitive system.

From the point of view of a biological interpretation, we observe that the mean frequency of an ensemble of responses to an isolated pulse coincides with the external frequency. So we suggest that, if the amoeba can be represented by this kind of frequency learning circuit, after a single pulse it responds, on average, with the same frequency of the train pulse with a certain (large enough) variance.

\begin{figure*}
\centerline{
\includegraphics[width=2\columnwidth]{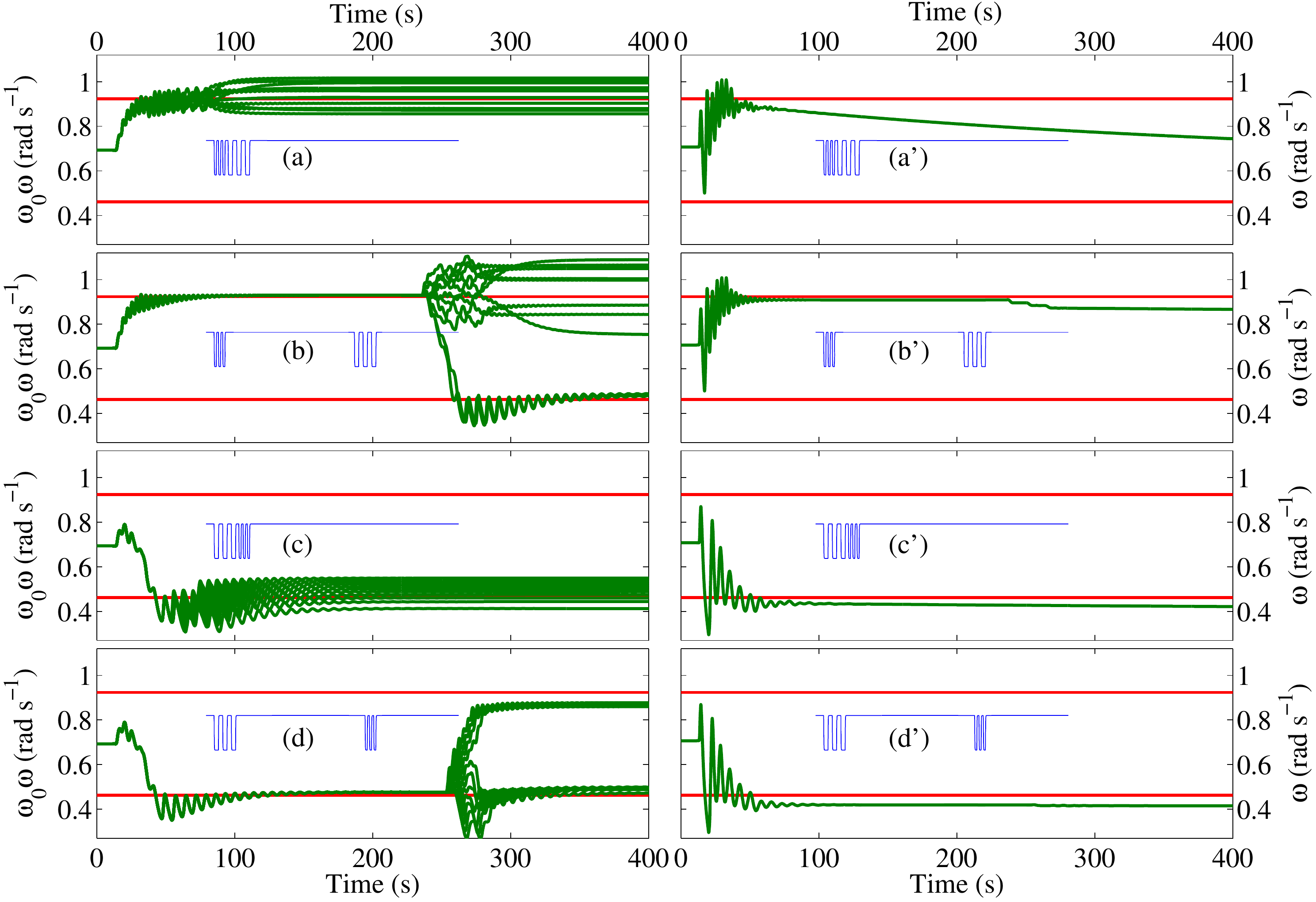}}
\caption{\label{Hopf_PDV_comparison2}Simulation of the LC contour from \insref{Circuits}{c}. While the left side presents the Hopf oscillator memcapacitive system results, the right side reports results of the DPT memcapacitive system model. The parameters are the same as in \figref{Hopf_PDV_comparison}. The insets represent the  applied voltage at $T_1=6.8\textnormal{s}$, $T_2=13.6\textnormal{s}$ and $V_P=2.7\textnormal{V}$. The thick red lines indicate the two pulse frequencies. }
\end{figure*}

Finally, in \figref{Hopf_PDV_comparison2}, the response to two different pulse trains is reported. When the trains are consecutive the first frequency is always the dominant. The fluctuations are due to the already discussed \Eqref{Hopf4} where, depending on the starting time of the second pulse train, the term depending on $i$ can be positive or negative and gives rise to different fluctuations of the response. On the other hand, as can be seen in \figref{Hopf_PDV_comparison2}, when the pulse trains are well separated, the first train switches off the memristive sytem, and as a consequence the second train has only a few chances to switch the system to the frequency of the second pulse train.

%%%%%%%%%%%%%%%%%%%%%%%%%%%%%%%%%%%%%%%%%%%%%%%%%%%%%%%%%%%%%
\subsection{DPT memcapacitive system}
%%%%%%%%%%%%%%%%%%%%%%%%%%%%%%%%%%%%%%%%%%%%%%%%%%%%%%%%%%%%%

We present here a different model of memcapacitive system that we name DPT (Di Ventra-Pershin-Traversa) model. This model describes a memcapacitive system capable of modifying its capacitance according to the external signal, and reach a constant steady state for periodic input signals. The adaptable LC-contour can be built using the circuit of \insref{Circuits}{c} and an effective third-order current-controlled memcapacitive system of memcapacitance $C_{eff}$
given by
\begin{equation}
C_{eff}  =\frac{c_1+\beta_1 x_3}{c_2+\beta_2^2\beta_3 L x_1}  \;\;\;.   \label{DVPT_C}
\end{equation}
The equations of motion for the internal state variables $x_i$ are selected as
\begin{eqnarray}
\dot x_1  &=& -\alpha_1 x_1+\beta_1 i^2 \label{DVPTeq_x1} \;\;\; ,\\
\dot x_2  &=& -\alpha_2 x_2+ \beta_2 i  \label{DVPTeq_x2} \;\;\; ,\\
\dot x_3  &=& -\alpha_3 x_3+\beta_3 x_2^2 \label{DVPTeq_x3} \;\;\;  .
\end{eqnarray}
Here, $\alpha_k$ are damping coefficients, and $c_k$ are used to define the state of the capacitance for a dc signal (zero current).

One can show that when driven by a sinusoidal voltage, the contour's frequency $\omega_{LC}=1/\sqrt{LC_{eff}}$ approaches $\omega_{ext}$ with time.
For the sake of simplicity, let's consider the DPT memcapacitive system driven by periodic current, $i=\tilde \imath\exp{\left(j\omega_{ext}t\right)}$ with $j$ the imaginary unit. In this case, in the frequency domain Eqs. (\ref{DVPTeq_x1})-(\ref{DVPTeq_x3}) read
\begin{eqnarray}
\tilde x_1  &=& \frac{\beta_1 \tilde \imath^2}{\alpha_1+j2\omega_{ext}} \label{DVPTeq_x1f} \;\;\; ,\\
\tilde x_2  &=& \frac{\beta_2 \tilde \imath}{\alpha_2+j\omega_{ext}}  \label{DVPTeq_x2f} \;\;\; ,\\
\tilde x_3  &=&\frac{ \beta_3 \beta_2^2 \tilde \imath^2}{(\alpha_2+j\omega_{ext})^2(\alpha_3+j2\omega_{ext})} \label{DVPTeq_x3f} \;\;\; .
\end{eqnarray}
Considering $c_1\ll\beta_1 x_3$, $c_2\ll\beta_2^2\beta_3 L x_1$ and $\alpha_i  \ll \omega_{ext}$, and substituting \Eqref{DVPTeq_x1f} and \Eqref{DVPTeq_x3f} into \Eqref{DVPT_C} we find $C_{eff}=-1/L\omega_{ext}^2$. The minus sign arises from the fact that $i$ is complex. Repeating the same calculation for $i=\tilde \imath(\exp{(j\omega_{ext}t)}+\exp{(-j\omega_{ext}t)})$ we obtain $C_{eff}=1/L\omega_{ext}^2$.

We tested the stability of this circuit with the same procedure described in section \ref{Hopf} and no evidence of instability regions were found. \figref{Hopf_PDV_comparison} reports the response to a pulse train plus an isolated pulse. The LC frequency approaches the external frequency (black solid line) as fast as the Hopf oscillator. However, it does not show any fluctuation as a function of the starting point of the pulses. This is simply due to the form of the equations.
In fact, when a constant voltage is applied to the memcapacitor for a long time, the variables $x_\alpha$ tend to $0$ (see \Eqref{DVPTeq_x1}-\eqref{DVPTeq_x3}). After that, when a time-dependent input is applied, the variables $x_\alpha$ grow fast, reaching their limit amplitude values given by \Eqref{DVPTeq_x1f}-\eqref{DVPTeq_x3f}. On the contrary, if the variables $x_\alpha$ have already the limit amplitude values given by \Eqref{DVPTeq_x1f}-\eqref{DVPTeq_x3f}, it takes much longer time to induce large relative variations of $x_\alpha$ by appropriate input signals.
Thus, when the system learns the frequency, the isolated pulse does not have any apparent effect on the LC frequency (see \figref{Hopf_PDV_comparison}).

The right column of \figref{Hopf_PDV_comparison2} shows the contour frequency response to a sequence of two pulse trains at different frequencies. When the second sequence immediately follows the first one the behavior is asymmetric. If the first pulse train is at higher frequency (HF) (\insref{Hopf_PDV_comparison2}{a\rq{}}) then the contour frequency initially converges to HF, but, when the second train with lower frequency (LF) finishes, the contour frequency begins to decrease reaching a minimum far from the HF.
This happens because the variables $x_\alpha$ have not enough time to reach the limit value given by \Eqref{DVPTeq_x1f}-\eqref{DVPTeq_x3f}. Correspondingly, the second pulse train is able to perturb the evolution of $x_\alpha$ and the final resonant frequency is in between HF and LF.
On the other hand, when the first pulse train is at LF (\insref{Hopf_PDV_comparison2}{c\rq{}}),
the first train is longer and the variables $x_\alpha$ have enough time to reach the limit values given by \Eqref{DVPTeq_x1f}-\eqref{DVPTeq_x3f}. Therefore,
the contour frequency converges to the LF and remains quite unperturbed. On the other hand, when the pulse trains are well separated (\insref{Hopf_PDV_comparison2}{b\rq{}} and (d\rq{})) the contour frequency remains quite unperturbed (just some small shifts are detectable when the first pulse train is at HF, see \insref{Hopf_PDV_comparison2}{b\rq{}}) and the system is able to adapt to only the first frequency.

We note that a possible way to discriminate between the Hopf oscillator and the DPT model, as the one
closest to the amoeba\rq{}s response, is to perform the two-frequency-experiment suggested in \figref{Hopf_PDV_comparison2}. In particular, if the pulse trains are well separated, the Hopf oscillator includes the possibility of learning (randomly) both LF and HF frequencies. On the other hand for the DPT model, under the same experimental conditions, only the first frequency can be learned.

%\subsection{Model 3}

%Finally, adaptive behavior can be based on stochastic processes. One such opportunity is offered by the stochastic gradient ascent algorithm that finds local maxima of given functions. The amplitude of driven oscillations $A(\omega,\omega_0)$ can be taken as the function seeking maximization (as a function of $\omega_0$). The maximum of $A(\omega,\omega_0)$ corresponds to the resonance condition $\omega_0 \simeq \omega$. Thus finding the maximum of $A(\omega,\omega_0)$  and adjusting the contour's frequency to $\omega_0$ we adapt the contour's frequency to the frequency of input oscillations.

%The equation of motion for $\omega_0$ can be written in the form
%\begin{equation}
%\frac{\textnormal{d}\omega_0}{\textnormal{d}t}=\beta A(\omega,\omega_0)\xi \;\;,
%\end{equation}
%where $\beta$ is a coefficient and the stochastic process $\xi$ is Gaussian. The memcapacitance $C$ is related to the $\omega_0$ through the usual relation $C=(L\omega_0)^{-1}$. $A(\omega,\omega_0)$ can be obtained introducing an internal state variable similarly to Eq. (\ref{eq_x1})...

%%%%%%%%%%%%%%%%%%%%%%%%%%%%%%%%%%%%%%%%%%%%%%%%%%%%%%%%%%%%%
%%%%%%%%%%%%%%%%%%%%%%%%%%%%%%%%%%%%%%%%%%%%%%%%%%%%%%%%%%%%%
\section{Summary} \label{summary}
%%%%%%%%%%%%%%%%%%%%%%%%%%%%%%%%%%%%%%%%%%%%%%%%%%%%%%%%%%%%%
%%%%%%%%%%%%%%%%%%%%%%%%%%%%%%%%%%%%%%%%%%%%%%%%%%%%%%%%%%%%%

The models of learning circuits proposed in this work suggest possible responses of simple biological organisms (that need to be experimentally verified) as well as new approaches for bio-inspired circuits. This section provides a  summary of our results in the form of comparison of the proposed circuits with learning/adaptive capabilities.

First of all, we note that learning and adaptability of biological organisms are very close features that, sometimes, are difficult to separate. By learning, we understand the ability to remember about the past. By adaptability, we understand behavioral changes based on past experiences. Clearly, the adaptability is not possible without learning. Moreover, there should be a tendency to loose capabilities that have been acquired by learning if they are not used 
for sufficiently long time.

One can make an attempt to separate learning and adaptability in our circuit models. Clearly, in all circuits in Fig. \ref{Circuits}, the states of memristive devices change when the circuit learns. However, the circuit in Fig. \ref{Circuits}(c) has an additional capability to adjust its frequency, so it memorizes both frequency and the fact of being trained. In this sense, Fig. \ref{Circuits}(c) circuit is ``smarter'' then the one in Fig. \ref{Circuits}(a) and (b), and represents more advanced biological organisms. However, we can not say, for example, that Fig. \ref{Circuits}(b) circuit does not learn the frequency. It learns it, but in a more simple way based on a set of contours of different frequencies.

As discussed in section \ref{sec:tr}, the learning threshold test reveals similar response of all models when no variation of external frequency is involved.  However,  the models show quite a different behavior if the input frequency changes. In the case of the multiple LC contours of Fig. \ref{Circuits}(b), a change in the input frequency activates a different LC contour keeping previously activated contours active. Consequently, the response of the entire system has a multiple frequency spectrum. In the case of Fig. \ref{Circuits}(c) with memcapacitive system based on the Hopf oscillator, the internal memcapacitor oscillator locks to the external frequency. In this case, the system is characterized by only one resonant frequency that can  adapt to the external one, and the adaptability involves the memory features of the memcapacitor. The last case we have considered is the LC contour employing the DPT memcapacitor. A distinctive feature of this contour is the speed of learning. The learning is fast from a blanc state and slow if the systems was trained in the recent past.

%%%%%%%%%%%%%%%%%%%%%%%%%%%%%%%%%%%%%%%%%%%%%%%%%%%%%%%%%%%%%
%%%%%%%%%%%%%%%%%%%%%%%%%%%%%%%%%%%%%%%%%%%%%%%%%%%%%%%%%%%%%
\section{Conclusions} \label{sec5}
%%%%%%%%%%%%%%%%%%%%%%%%%%%%%%%%%%%%%%%%%%%%%%%%%%%%%%%%%%%%%
%%%%%%%%%%%%%%%%%%%%%%%%%%%%%%%%%%%%%%%%%%%%%%%%%%%%%%%%%%%%%

In conclusion, we have discussed several models of memory that are both inspired by the primitive learning abilities of unicellular organisms, as well as are meant to provide biological feedback through the proposal of several experiments that can distinguish between the different circuit responses. In particular, we have looked at the response of a set of LC-contours with memristive damping, and a single memcapacitive system-based adaptive contour with memristive damping. In the latter case, we have also suggested a new memcapacitive model that adapts to the frequency of the input signal.

Overall, both circuits from Figs. \ref{Circuits}(b) and \ref{Circuits}(c) satisfy all {\bf F}-s and {\bf A}-s mentioned in the Introduction ($\bf A2$ and $\bf A3$ are achieved by the use of Eq. (\ref{Vcontr2})). Moreover, both circuits satisfy $\bf S1$ postulating the learning threshold that still needs to be verified experimentally. While Fig. \ref{Circuits}(b) circuit satisfies $\bf S2$, Fig. \ref{Circuits}(c) circuit does not satisfy this criterion. An experimental study of the response of the  biological system to two frequency pulses needs to be performed to discriminate between the models. Clearly, the progress in understanding the memory mechanisms in simple organisms (including its threshold switching properties) depends critically on future experiments with real organisms.

In addition to serve as models of biological processes, the considered circuits are clear examples of adaptive {\it passive} electronics \cite{driscoll2010memristive}. They may then find application in several areas of technology, e.g., signal processing, pattern recognition, and even unconventional computing.

\section{Acknowledgment}

This work has been partially supported by the Spanish government through project TEC2011-14253-E and by NSF grants No. DMR-0802830 and ECCS-1202383, and the Center for Magnetic Recording Research at UCSD.

%\section*{References}

\bibliographystyle{ieeetr}
\bibliography{bibliogrphy_tnnls}

\end{document}